\documentstyle[11pt,newpasp,twoside,epsf]{article}
\reversemarginpar

\begin{document}

\title{AD Leo from X-Rays to Radio:
      Are Flares Responsible for the Heating of Stellar Coronae?}

\author{M. G\"udel, M. Audard}
\affil{Paul Scherrer Institut, W\"urenlingen and Villigen, CH-5232 Villigen PSI,
       Switzerland}
\author{Edward F. Guinan}
\affil{Dept. of Astronomy \& Astrophysics, Villanova University, Villanova, PA 19085, USA}
\author{Jeremy~J. Drake, Vinay L. Kashyap}
\affil{Harvard-Smithsonian Center for Astrophysics, Cambridge, MA 02138, USA}
\author{Rolf Mewe}
\affil{SRON, Sorbonnelaan 2, 3584 CA Utrecht, The Netherlands}
\author{I.~Y. Alekseev}
\affil{Crimean Astrophysical Observatory, Nauchny, Crimea 334413, Ukraine}

\begin{abstract}
In spring 1999, a long coordinated observing campaign was performed on the flare 
star AD Leo,
including {\it EUVE}, {\it BeppoSAX}, the {\it VLA}, and optical telescopes.  
The campaign covered a total of 44 days. We obtained high-quality light curves
displaying ongoing variability on various timescales, raising interesting 
questions on the role of flare-like events for coronal heating. 
We performed Kolmogorov-Smirnov tests to compare the observations with a large
set of simulated light curves composed of statistical flares that are distributed  
in energy as a power law of the form ${\rm d}N/{\rm d}E \propto E^{-\alpha}$
with selectable index $\alpha$. We find best-fit $\alpha$ values slightly above 
a value of 2, indicating that the extension of the flare population toward small 
energies could be important for the generation of the   overall X-ray emission. 
\end{abstract}

\section{Introduction}

While large, episodic stellar flares heat coronal plasma rather efficiently
up to 100 MK for minutes to hours, it is the potentially large number
of small, and even `undetected'  flares (``microflares'', ``nanoflares'', 
e.g., Parker 1988) that have recently attracted the attention of both solar and 
stellar research.  Solar observations show that the flare occurrence rate is 
distributed in  energy as a power law, 
\begin{equation}
{{\rm d}N\over {\rm d}E} \propto E^{-\alpha} 
\end{equation}
(e.g., Lin et al. 1984). The value of $\alpha$ determines the importance of 
the low- or high-energetic tail of the distribution: The integration of the energy
over the energy distribution (1)  
\begin{equation}
L_{\rm X} \propto  \int_{\rm E_{\rm min}}^{\rm E_{\rm max}} 
{{\rm d}N\over {\rm d}E}E {\rm d}E 
\end{equation}
with $\alpha \ge 2$ produces arbitrarily large total emission rates
if $E_{\rm min} \rightarrow 0$ (microflares, nanoflares).
There is ample evidence that the ensemble of flares play a fundamental role
in the heating of (quiescent) coronae of magnetically active stars:
\begin{itemize}

\item Active stars emit strong gyrosynchrotron radio emission during 
     quiescence: Evidence for accelerated (MeV) electrons as in solar flares (G\"udel 1994).
     
\item This non-thermal emission correlates with the overall X-ray radiation 
      the same way as   solar flares do (G\"udel \& Benz 1993; Benz \& G\"udel 1994).
      
\item The optical {\it U} band  flare frequency correlates with the quiescent 
      X-ray stellar luminosity (Doyle \& Butler 1985).
      
\item Transition region lines show broadening probably related to explosive events 
     (Wood et al. 1996).               

\item Small flare events with energies of the order
     of $10^{27} - 10^{28}$~ergs have become observable with the Hubble Space
      Telescope in cool M dwarfs, proving their ubiquity in active stellar atmospheres 
     (Robinson et al. 1995).

\item The structuring of X-ray emitting active coronae is reminiscent
      of the thermal structure of flares (G\"udel et al. 1997) and may be
      explained by the superposition of a distribution of statistical 
      flares (Kopp \& Poletto 1993; G\"udel 1997).

\item Solar observations now show the importance of micro- and nanoflares in
      coronal energy release: $\alpha = 2.3-2.6$ for microflares in the quiet corona
      (Krucker \& Benz 1998).
      
\end{itemize}
Studies of statistical flare distributions  have been rare in the {\it stellar} context, 
due to the paucity of
relevant data sets. Collura et al. (1988) found a power-law index
$\alpha = 1.52$ from {\it EXOSAT} observations of dMe stars, while Osten \& Brown (1999)
report $\alpha = 1.6$ for a sample of RS CVn binary systems observed with {\it EUVE}.
In a series of papers (Audard, G\"udel, \& Guinan 1999; Audard et al. 2000; G\"udel
et al. 2000a) we have been investigating systematically the role of {\it statistical} 
flares in the overall coronal heating of active stars. From a large sample of {\it EUVE} 
observations, we  found that 
\begin{itemize}
\item relatively steep ($\alpha = 2-2.5$) power laws dominate the flare rate 
   distributions, indicating that {\it small flares are important} in these coronae.
   
\item Statistical heating by flares leads to dominant {\it coronal temperatures}
      in agreement with the measurements.
      
\end{itemize}

\section{New Observations of AD Leo: Flares over and over again}

An extremely long {\it EUVE} observation of AD Leo was approved during {\it EUVE}'s cycle 
7 between April 2 and May 16, 1999 (with  a few minor time gaps) comprising
 900 ksec of on-source exposure time. The DS instrument and the
three spectrometers (SW, MW, LW) were used.
Between May 1 and May 15, we obtained a total of 270~ks of exposure
time with {\it Beppo}SAX, covering about 8 days within this interval. We obtained
data from the LECS (0.1--10~keV), the MECS (2--10~keV), and the PDS
(15--400~keV) instruments.
On April 29, a 10~hr integration was carried out with the  {\it VLA}.
We used the 2~cm, 3.6~cm and 6~cm bands with the array in its D configuration.  
Two  optical photometry observatories (Villanova University and 
Crimean Astrophysical Observatory) observed AD Leo in the {\it U, B, V, R,} and
{\it I} bands during 
several nights in April and May.

\begin{figure}
\plotone{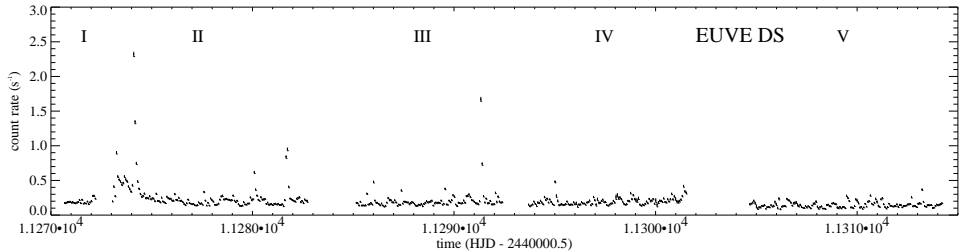}
\caption{{\it EUVE} DS light curve of AD Leo, obtained between April 2, 1999, and May 16, 1999.
          Segment V suffers from `dead spot' reduction in effective area and high radiation. The error bars
	  are typically $\pm$0.01~ct~s$^{-1}$ and have been plotted.}
\end{figure}

\begin{figure}
\plotone{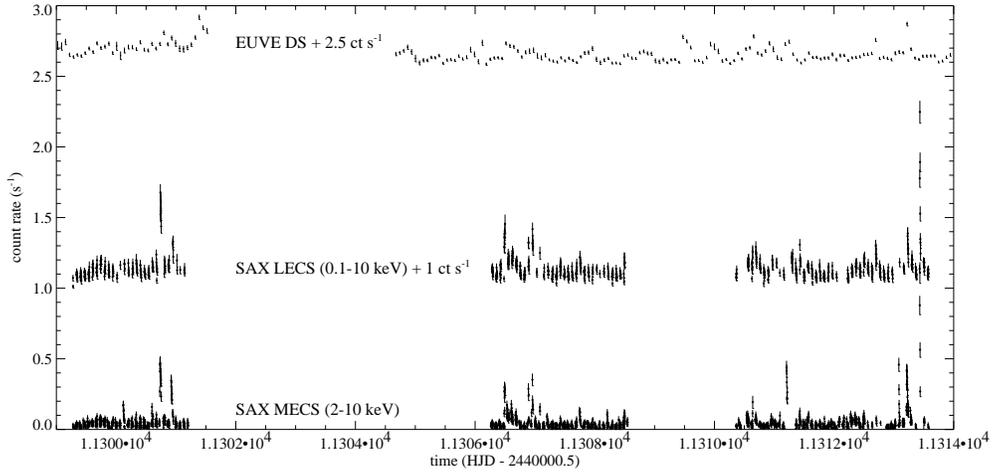}
\caption{{\it BeppoSAX} MECS and LECS light curves, compared with simultaneous
     {\it EUVE} DS data. For illustration purposes, the LECS and {\it EUVE} light curves have been
     shifted by +1 and +2.5 cts~s$^{-1}$.}
\end{figure}

AD Leo is a dM4.5e star with a rotation period of 2.7~d (according to our 
[EFG] measurements, $P_{\rm rot}$ appears to be 1.7~d only).
An overview of the {\it EUVE} and {\it Beppo}SAX observations is shown in Figures 1$-$2
(also G\"udel et al. 2000b).
The {\it EUVE} light curve is extremely variable. Most of the fluctuations visible
in Fig. 1 are real (the error bars being smaller than the visible fluctuations).

\section{Measuring the Importance of Flare Heating}

The light curves of AD Leo at hand are ideal for further investigation of statistical
flares. We address two questions: (i) What is the statistical distribution of the {\it visible}
flares in energy? (Given the lack of detailed spectroscopy for each flare,
we approximate the total energy by the total number of counts detected, times a constant
count-to-energy conversion factor.) (ii) Can the complete emission, including the
apparently quiescent radiation, be explained by a superposition of statistical
flares with a distribution that is compatible with (i)? We use two  approaches to address
these issues.

\subsection{Kolmogorov-Smirnov Statistical Tests of the Flux Distribution}

Here, we investigate the distribution  of count rate values in the light curve. We proceed
as follows:
\begin{description}
\item[(i)] We determine the average flare profile through autocorrelation analysis from the
     observations.

\item[(ii)] We simulate light curves that are composed of statistical flares. These are
      distributed in energy according to a power law (1).  We use 5480 bins, which 
      is about ten times the number of bins in the {\it EUVE} light curve  (if binned to one
      data point per satellite orbit), i.e., we simulate ten statistical realisations
      of our observation.
       
\item[(iii)] The simulation is renormalized to the observation. To this end,
     we determine the cumulative distribution of flux values both for the
     observation and for the simulations (i.e., number of
     bins with a flux exceeding a given flux - see Figures 3$-$4). 
     We normalize the model flux to the observed flux by adjusting the middle portion of 
     the cumulative distribution, which lies above the noise level but at fluxes that 
     are frequently attained.

\item[(iv)] Statistical noise corresponding to the observations is added.

\item[(v)] We then perform a Kolmogorov-Smirnov (KS)  statistical test between
     simulation and observation by comparing the number  distribution of flux values
     in the available bins. {\bf Option:} The dominant `quiescent' part which
     is difficult to model can be subtracted beforehand to avoid problems
     with overlapping flare wings, rotational
     modulation effects, some true long-term variability  etc. 

\end{description}
Figures 3$-$4 show best-fit realisations for different $\alpha$ values.
   The top panel shows the data used, the second row  refers to the optimum 
   $\alpha$ found, while the other panels
   show cases for too large and too small $\alpha$. In each case, the solution with
   the highest confidence was identified by varying the number of flares in the
   simulation. 

\newpage
{\ }
\begin{figure}[h!]
{\ }\vskip -0.8truecm
\plotone{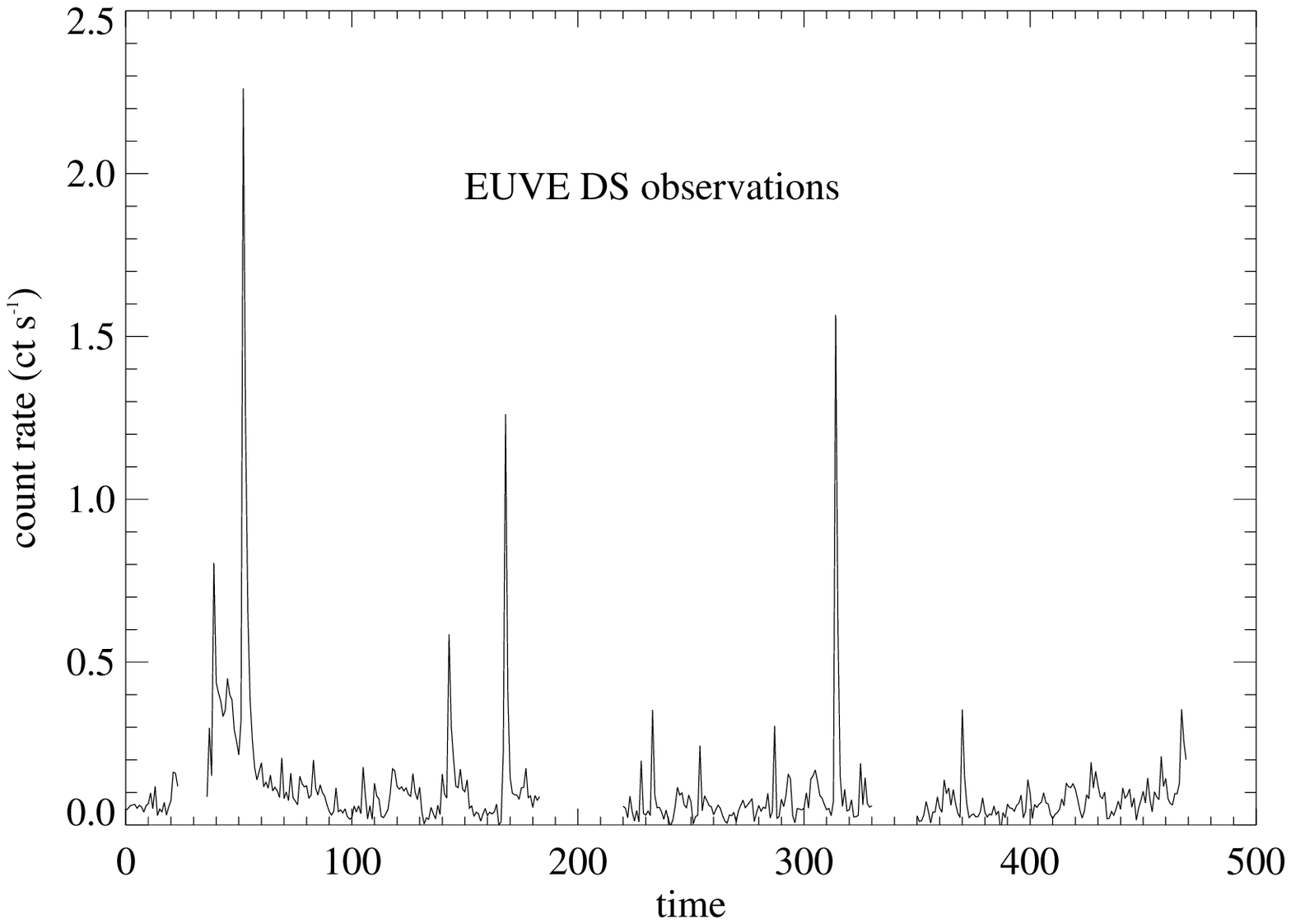}
\end{figure}
\nopagebreak
\begin{figure}[h!]
\vskip -1.0truecm
\plottwo{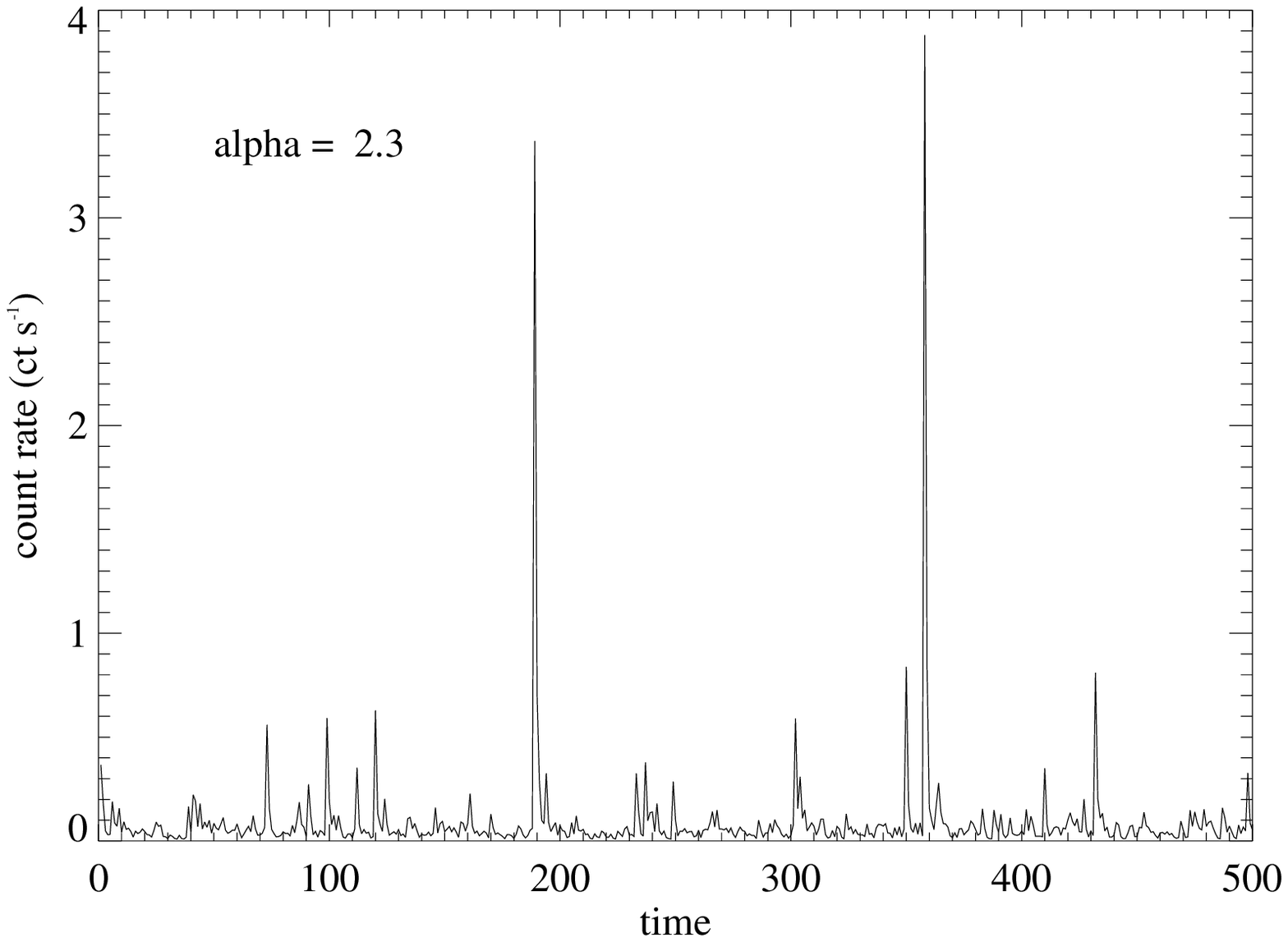}{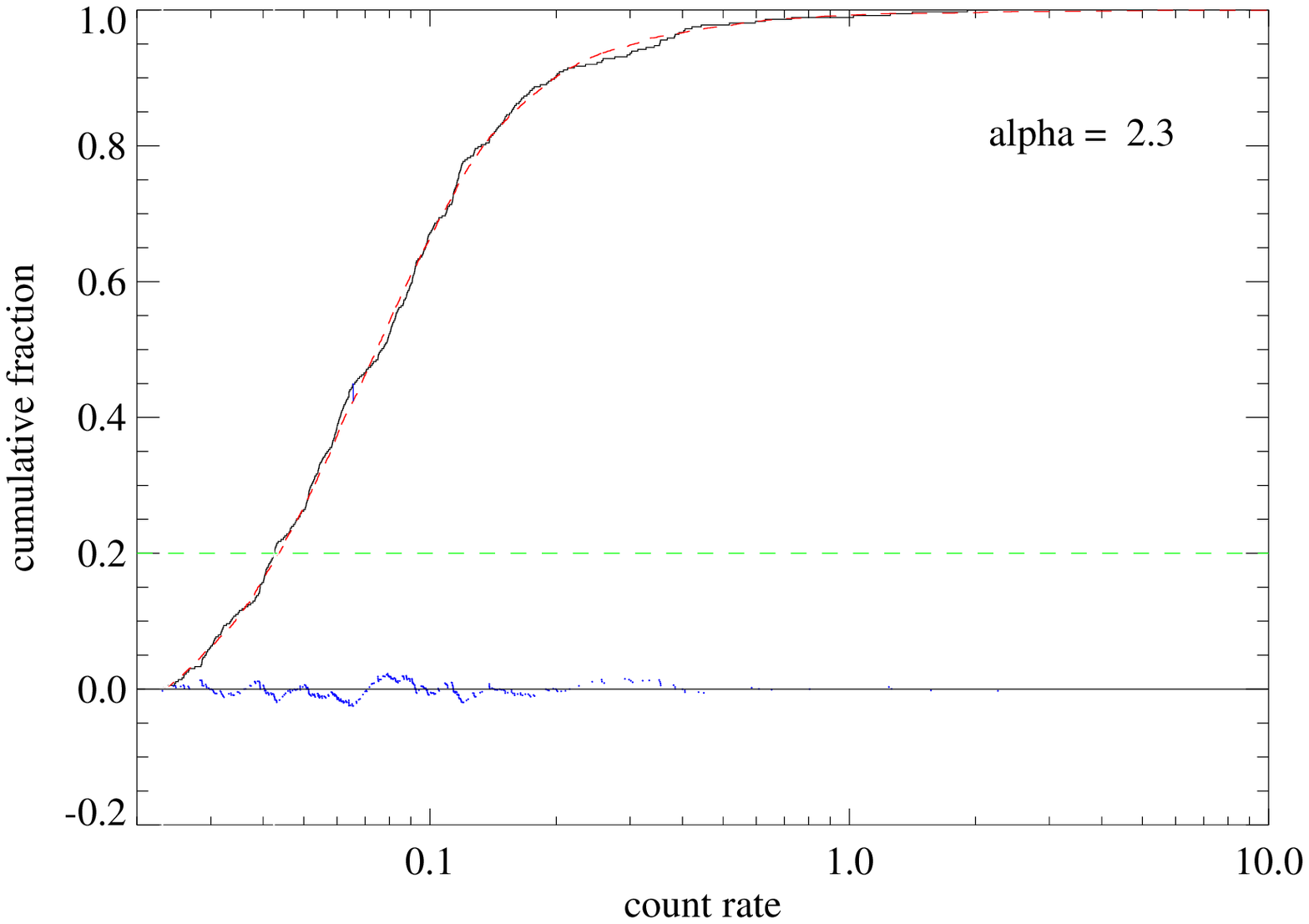}
\end{figure}
\nopagebreak
\begin{figure}[h!]
\vskip -1.0truecm
\plottwo{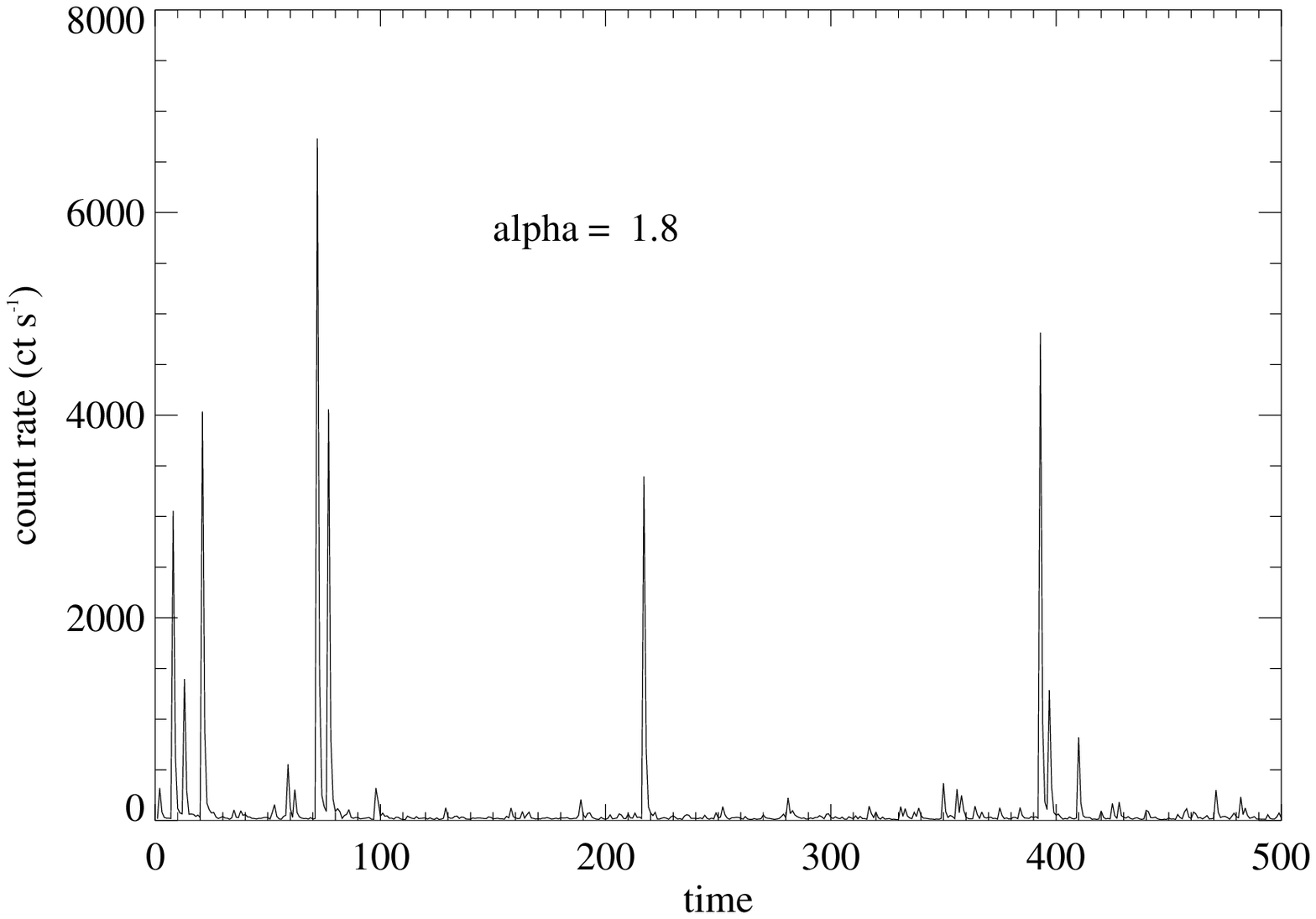}{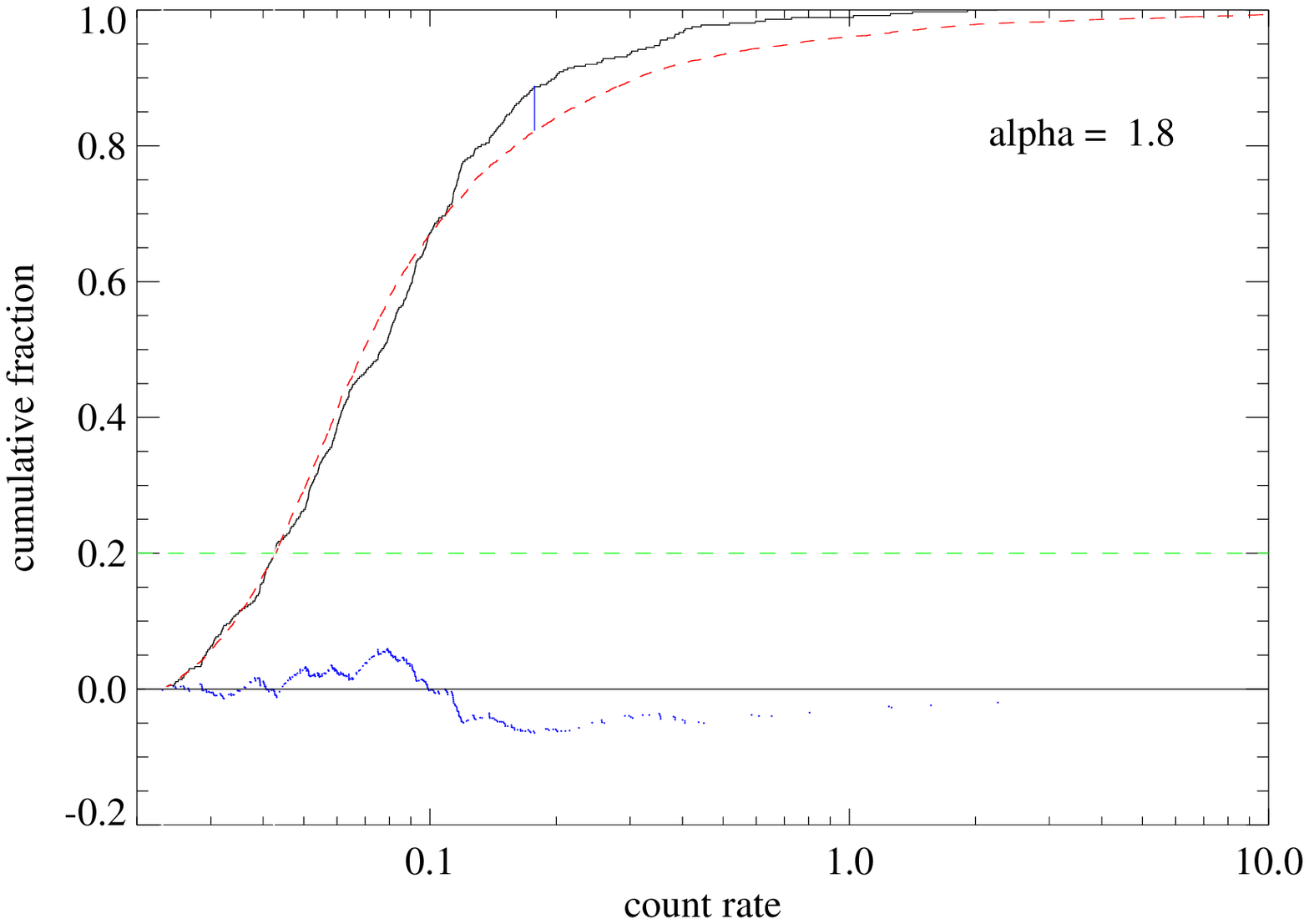}
\end{figure}
\nopagebreak
\begin{figure}[h!]
\vskip -1.0truecm
\plottwo{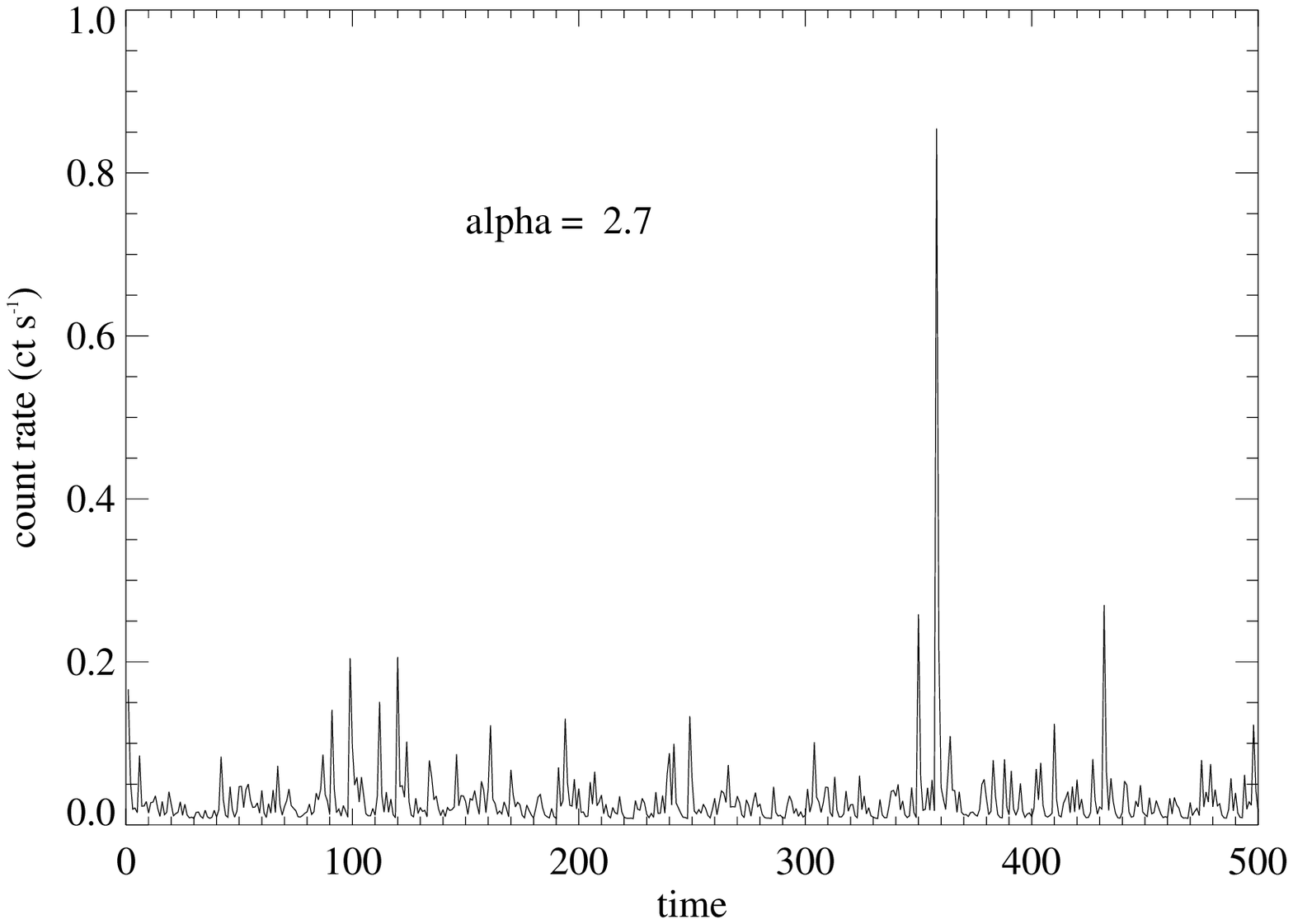}{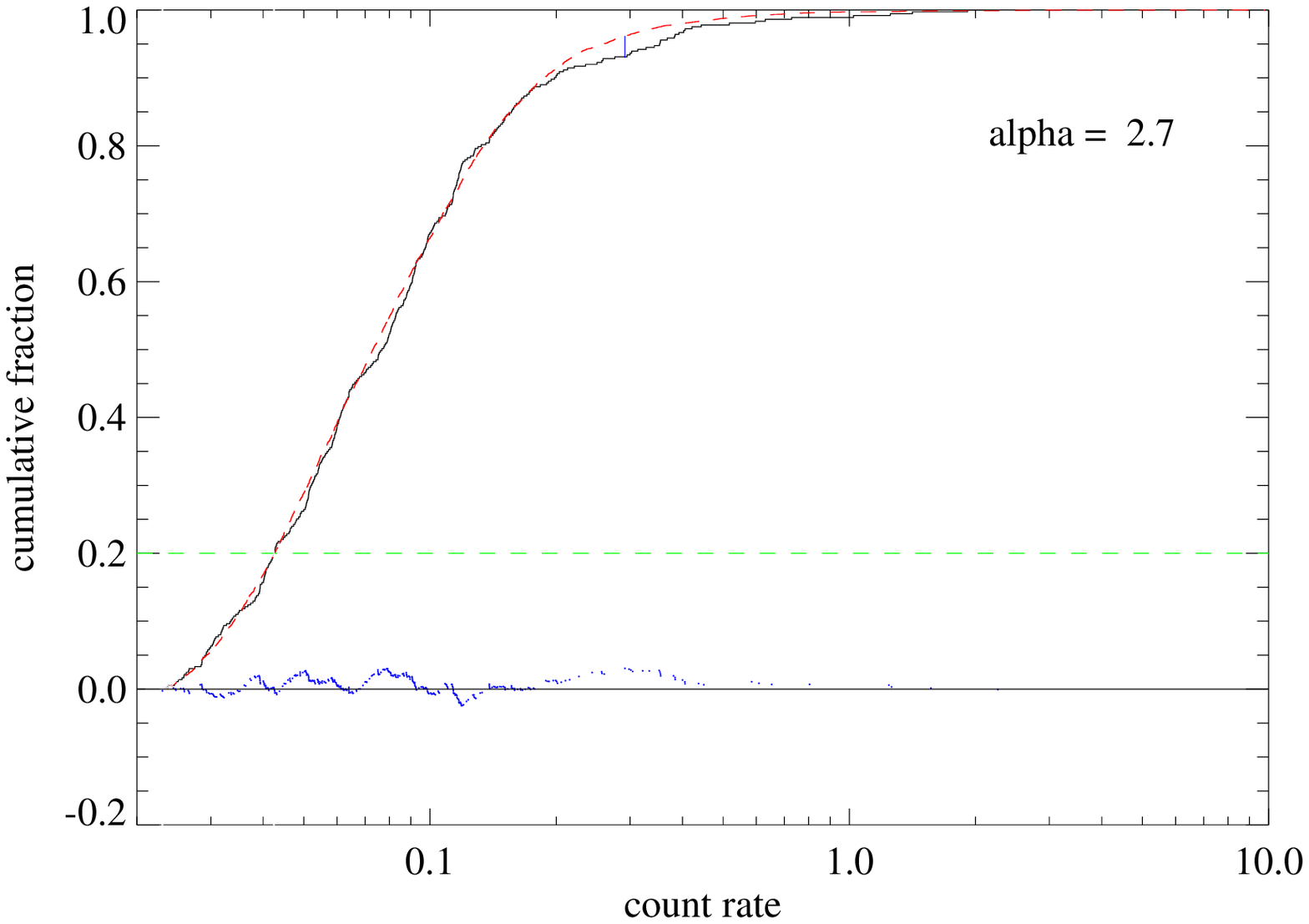}
\vskip -0.2truecm
\caption{Best-fit examples of statistical flare simulations for different power-law
         distributions. Only emission exceeding `quiescent' level has been modeled. 
	 {\bf Left:} Simulated light curve (extract, unnormalized). {\bf Right:}
	 Cumulative flux distribution for data (black) and model (red), and
	 difference (blue dotted). The maximum difference (blue vertical bar) 
	 is relevant for the KS statistic.
	 {\bf Top:} Optimum case; $\alpha = 2.3$ (statistical confidence: 98.2\%).
	    {\bf Middle:} Too hard distribution
	 with $\alpha = 1.8$ (statistical confidence for best case: 11.7\%).
	  {\bf Bottom:} Too soft distribution with $\alpha = 2.7$ (statistical 
	  confidence for best case: 84.9\%).}
\end{figure}

\begin{figure}
\plotone{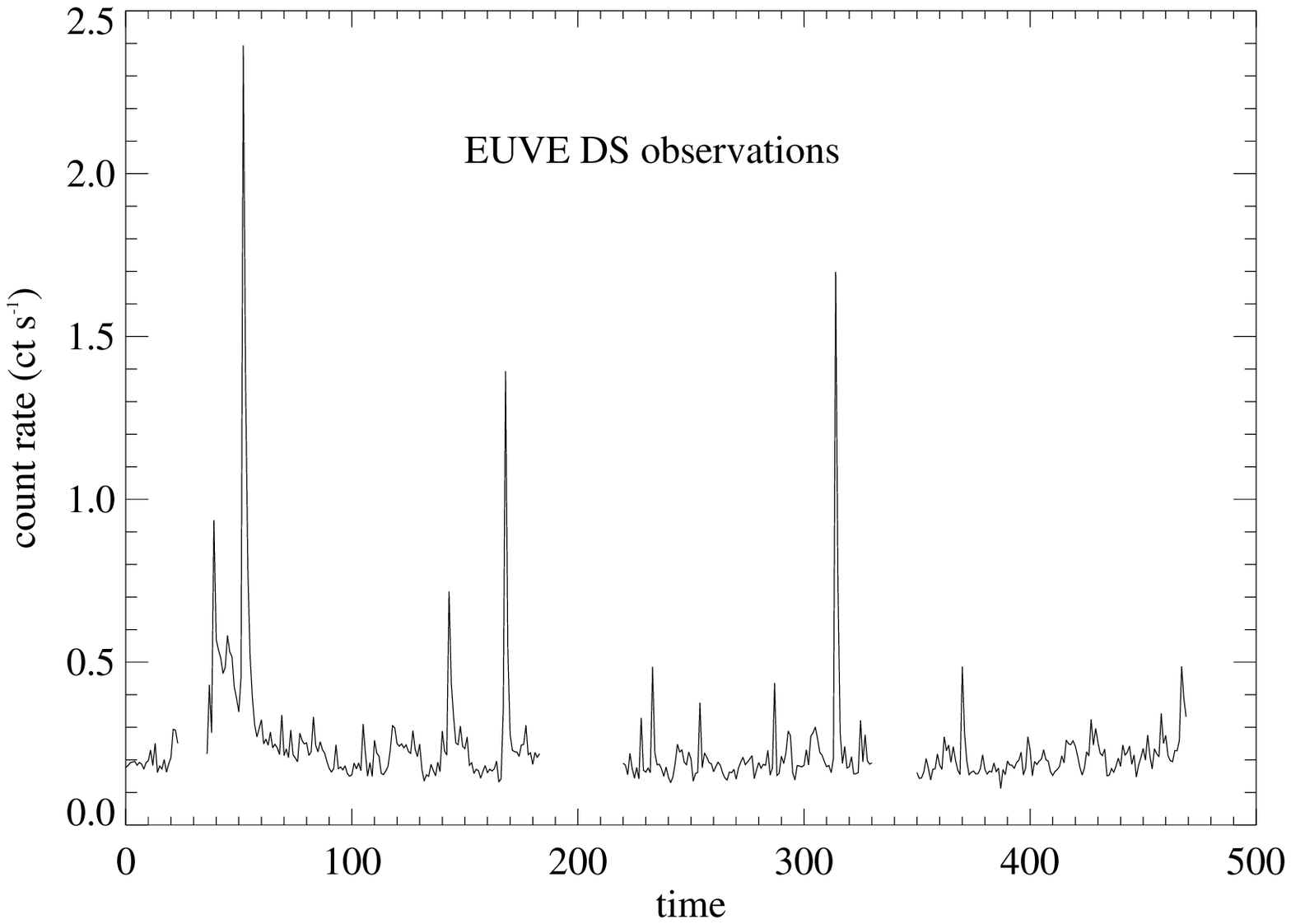} 
\end{figure}
\begin{figure}
\vskip -0.7truecm
\plottwo{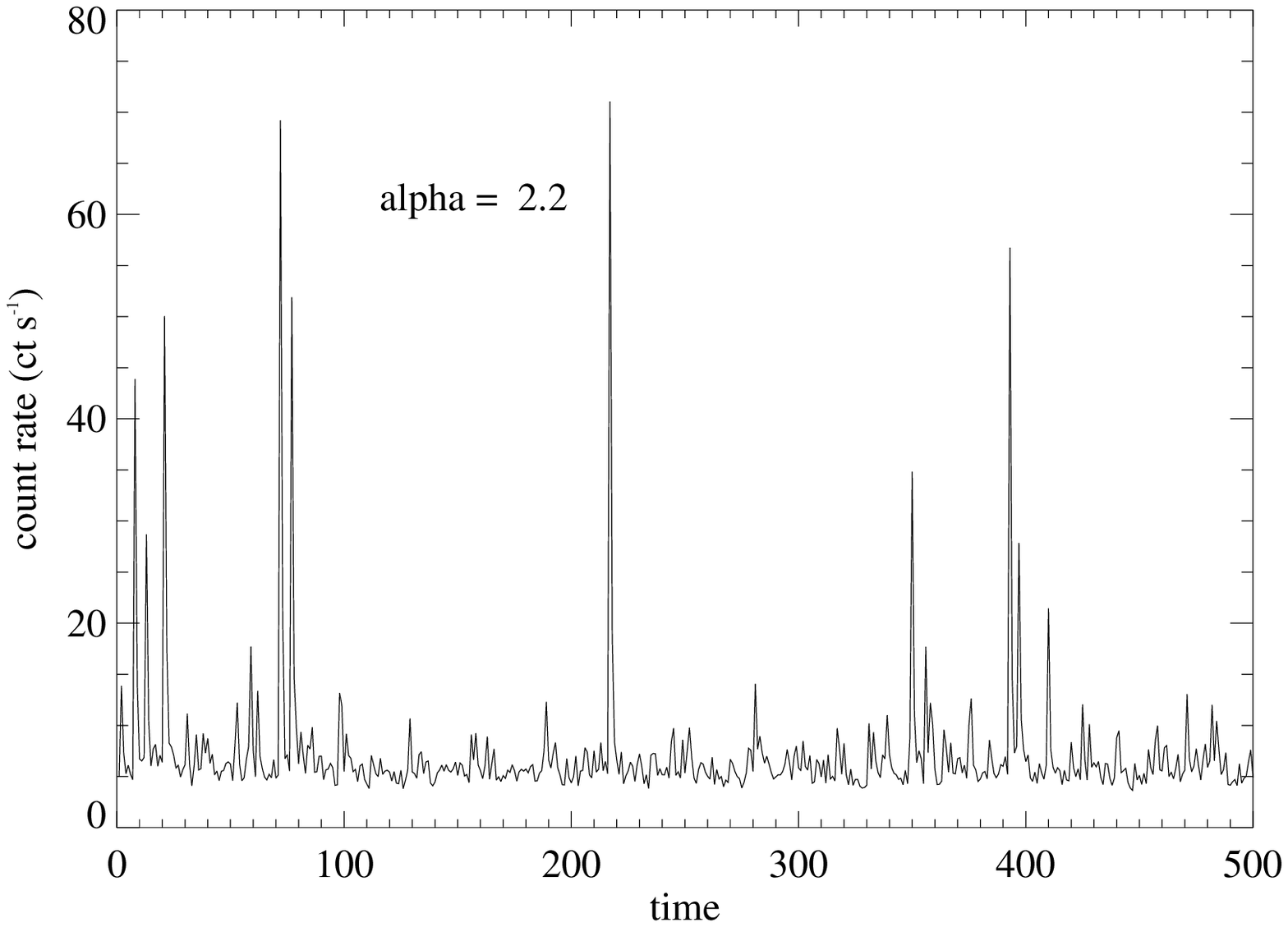}{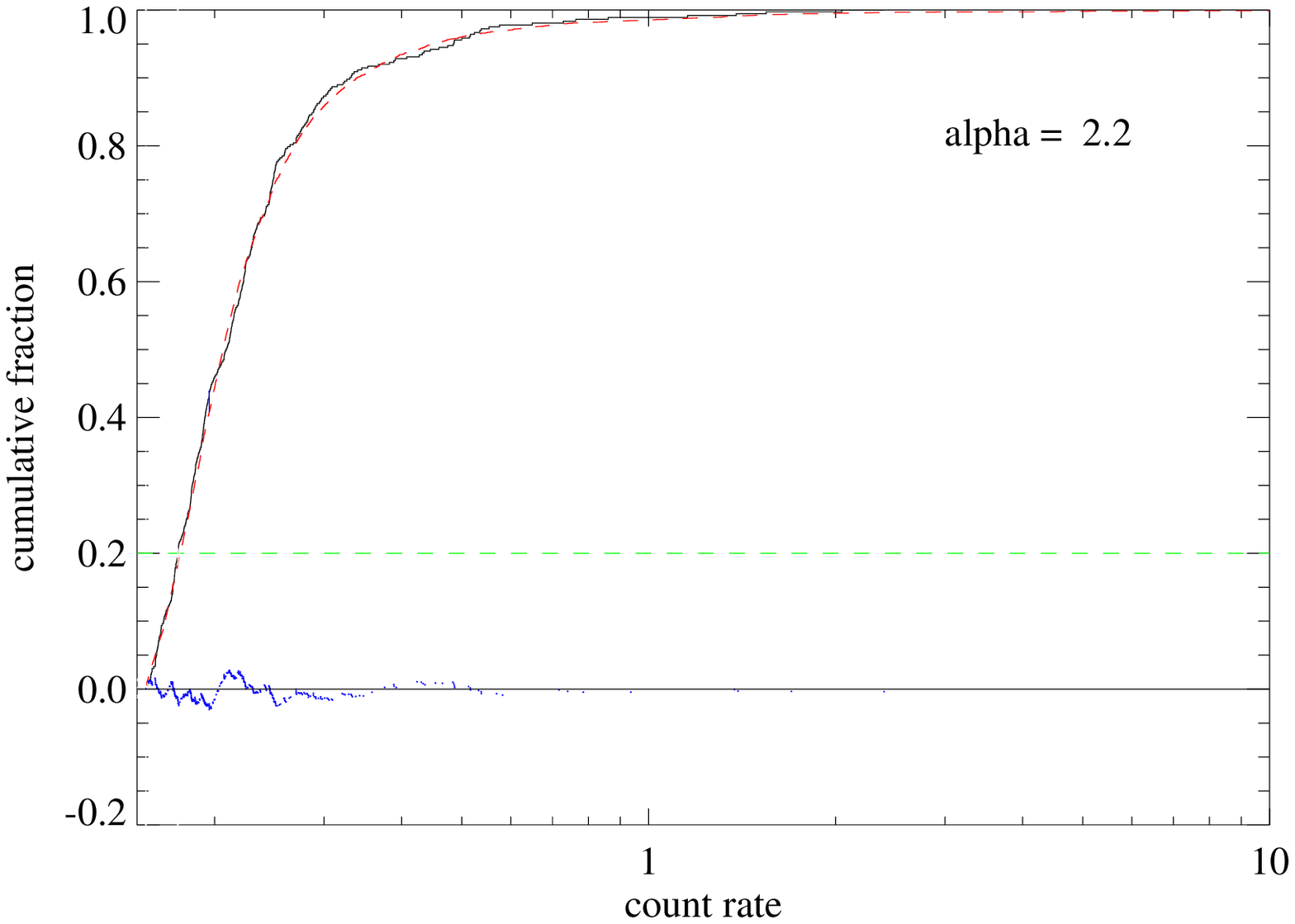}
\end{figure}
\begin{figure}
\vskip -0.7truecm
\plottwo{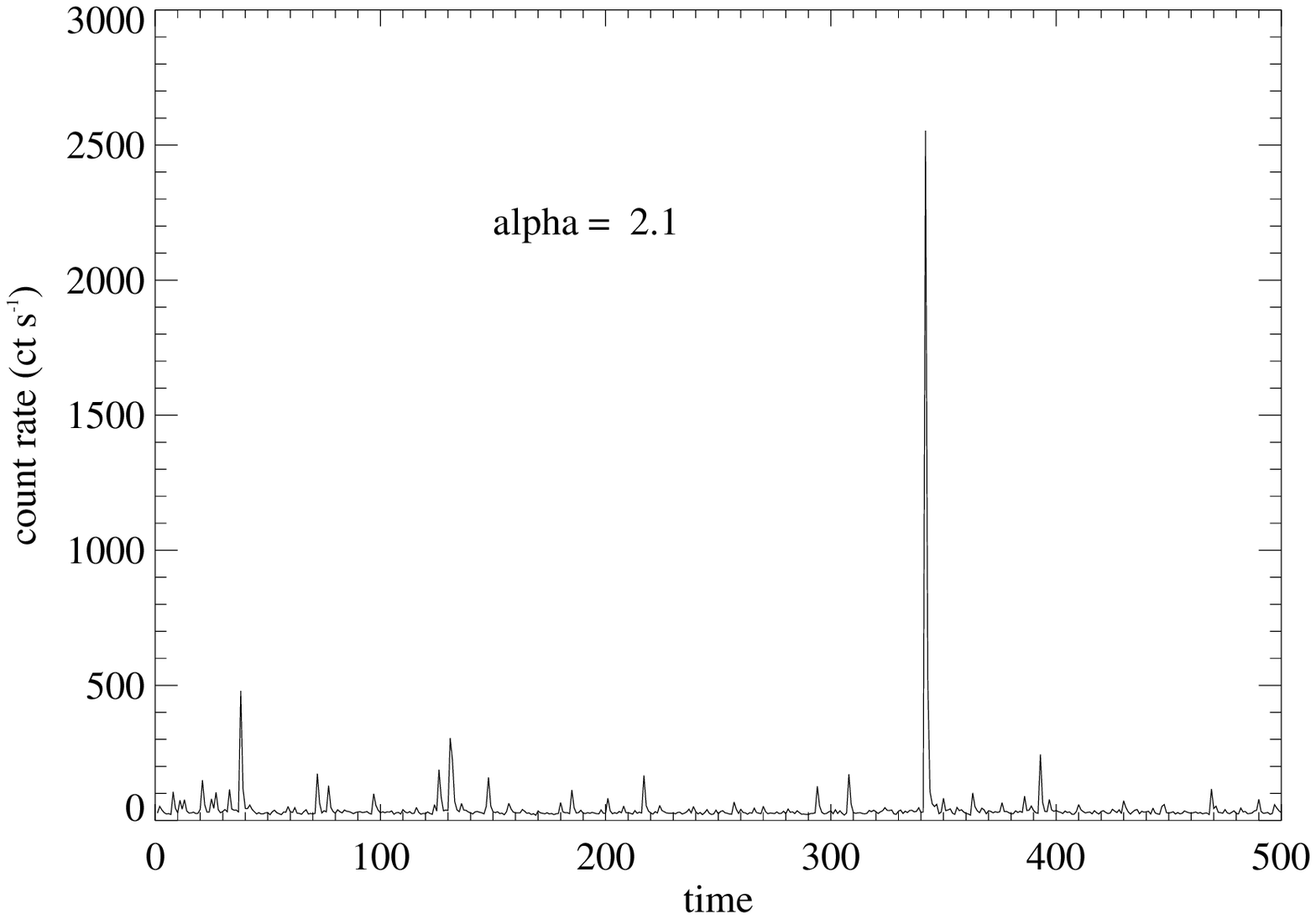}{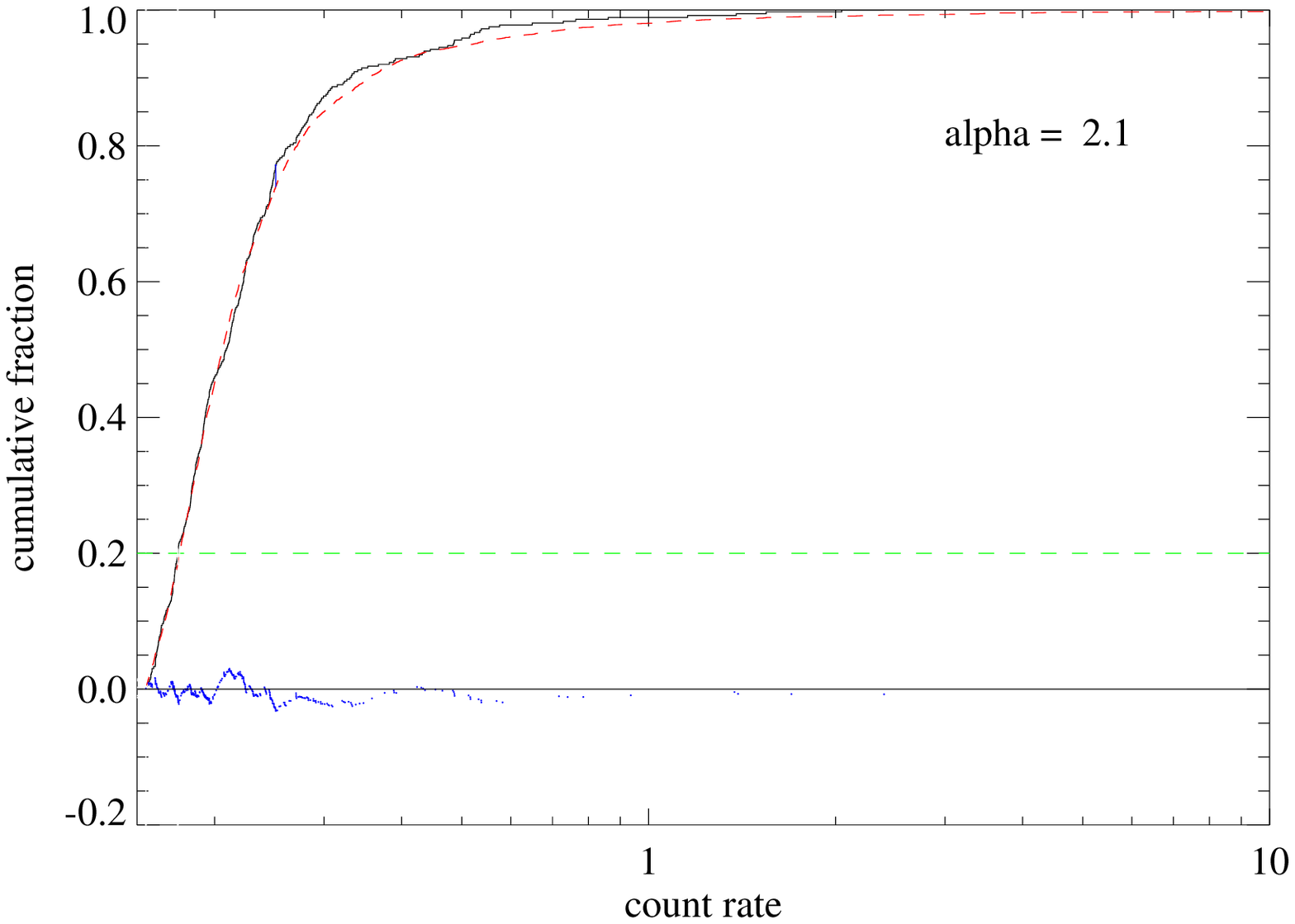}
\end{figure}
\begin{figure}
\vskip -0.7truecm
\plottwo{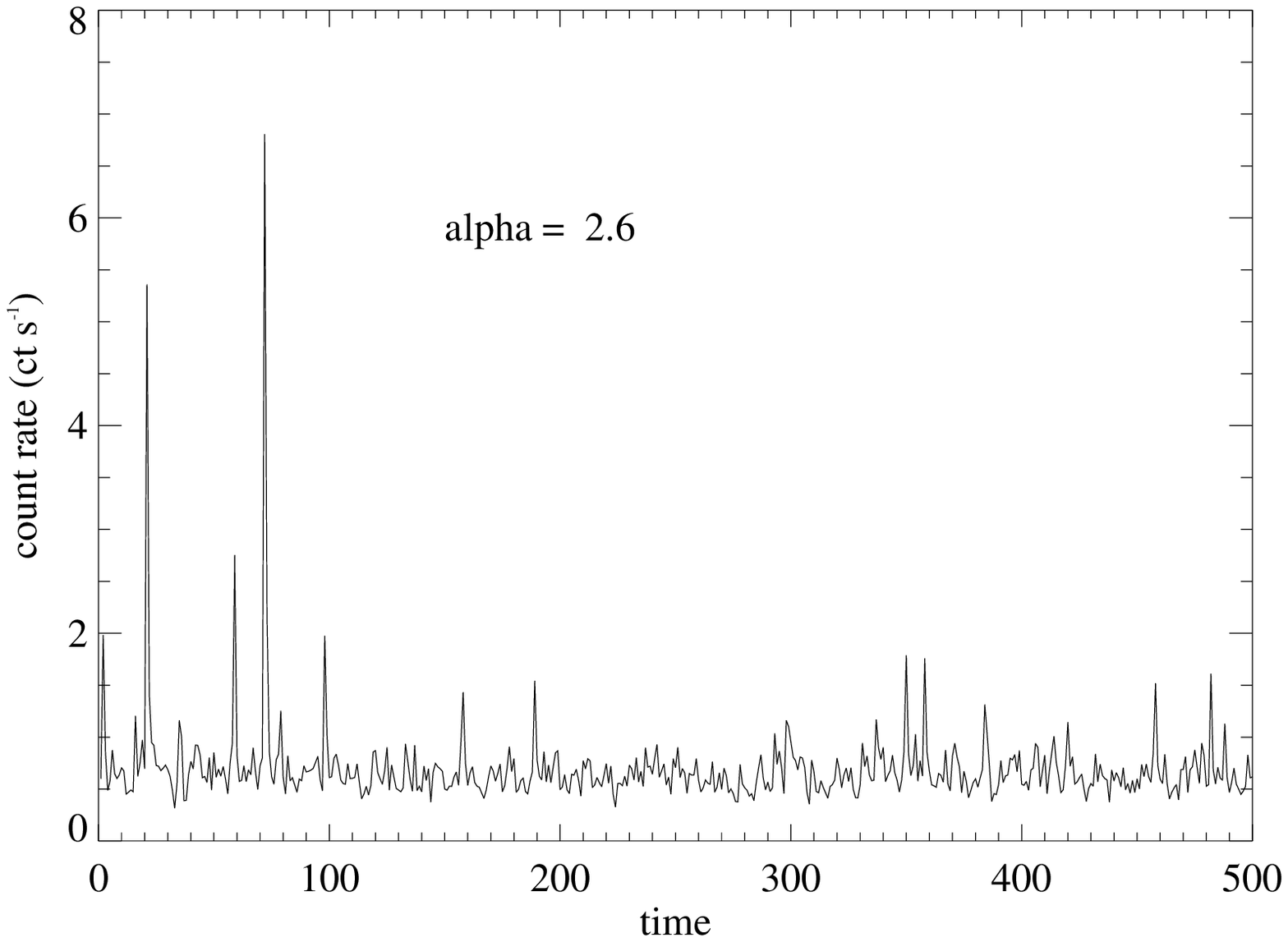}{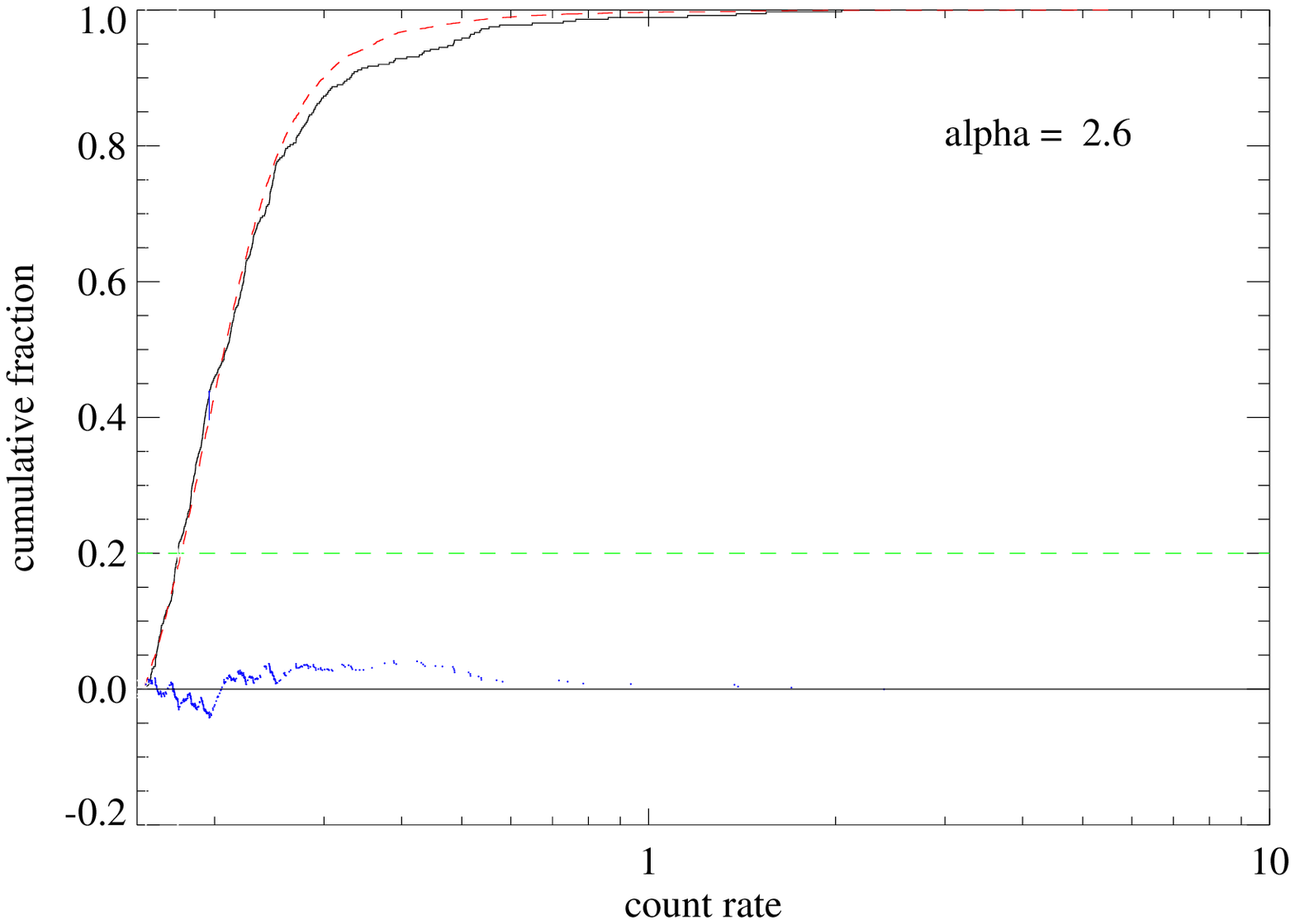}
\caption{Similar to Fig. 3, but all emission, including quiescent emission, has been modeled. 
	 {\bf Top:} Optimum case; $\alpha = 2.2$ (statistical confidence: 91.8\%).
	    {\bf Middle:} Too hard distribution
	 with $\alpha = 2.1$ (statistical confidence for best case: 86.6\%).
	  {\bf Bottom:} Too soft distribution with $\alpha = 2.6$ (statistical 
	  confidence for best case: 53.1\%).}
\end{figure}

\newpage
\noindent The left panels show the first 500 bins of the 
simulated light curves, while the right panels show the cumulative distributions
of count rates attained for the observation (black, solid) and the simulation (red, dashed).
We do not use the lowest 10\% of the distribution since those flux levels may be influenced
by slowly varying emission in the `quiescent' emission.
The vertical differences between the simulated and the observed
distribution are indicated dotted (blue) around the zero level: deviations to positive values
indicate a locally too soft simulated distribution, negative deviations indicate locally
too hard simulated distributions. The maximum deviation is indicated by a blue vertical
bar. The (green) dashed horizontal line indicates the lower threshold
for identifying the maximum vertical difference.

Since the last segment in the {\it EUVE} DS
light curve suffered from incursion into the DS dead spot and from elevated
particle radiation, we omitted that segment in our analysis. We plot in Fig. 
5 the best results for different $\alpha$ values, for  three cases:
(i) Modeling of the {\it EUVE} DS emission above the `quiescent' level (A);
(ii) modeling of the  complete {\it EUVE} DS emission (B); (iii)  modeling the complete 
{\it Beppo}SAX LECS light curve (C). Acceptable models ($>$90\% confidence) 
are found for  $\alpha = 2.2-2.5$.

\begin{figure}
\plotone{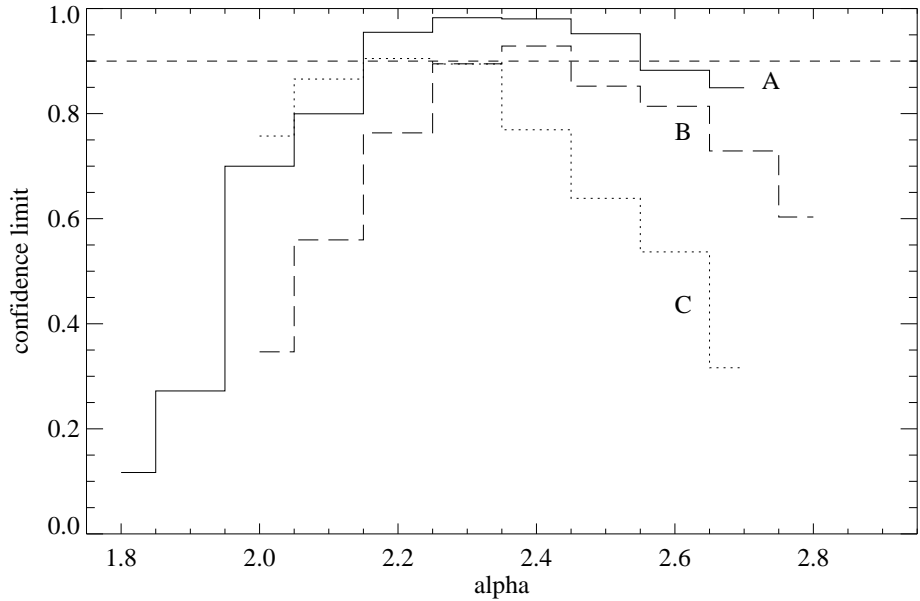}
\caption{KS confidence values for different $\alpha$, using those flare rates that produce
        the highest confidence values for a given $\alpha$ (for the hypothesis that the
	simulations and the observation are drawn from the same distribution).
         (A): {\it EUVE} DS, emission above quiescent
	level. (B): {\it EUVE} DS, all emission. (C): {\it Beppo}SAX LECS, all emission.}
\end{figure}

\subsection{Effect of Weak Flares}

The high time-resolution of the EUV events ($\sim 1 ms$) also affords
the possibility of detecting weaker flares in the light-curve (see
Kashyap et al.\ 2000).  By comparing the distribution of photon
arrival-time differences in the data stream with that expected from
a model of flare distributions with specific power-law indices, we
can account for the weak (but numerous) flares that contribute to the
emission.  Specifically, we
\begin{itemize}
\item[{\bf (i)}]
adopt a power-law flare-distribution model (Equation 1) described by
the power-law index $\alpha$ and the normalization $N_f$, and generate
a high-time-resolution light-curve via Monte-Carlo simulation;
\item[{\bf (ii)}]
add a constant component $N_b$ to the flare light-curve to account for
true steady emission, flare emission too weak to be distinguished
from steady emission, and background;
\item[{\bf (iii)}]
obtain a set of event times from the model light curve;
\item[{\bf (iv)}]
compare the observed distribution of arrival-time differences with that
derived from the generated model event-times and compute a test statistic
similar to $\chi^2_r$; and
\item[{\bf (v)}]
carry out the comparison over a grid of parameter values ($\alpha$, $N_f$,
$N_b$) to find the best-fit and confidence range.
\end{itemize}
We have run the above algorithm on the first half of segment {\tt II}
of the AD Leo EUVE observation (see Figure 1) and find that
$\alpha=2.1 \pm 0.05$, a smaller value than found above, but yet $>2$.
These data are clearly dominated by the large flare; an analysis of
longer segments, which is expected to reduce the bias due to this flare
and provide a more realistic assessment of the value of $\alpha$, is in
progress.
\section{Discussion and Conclusions}
         
The analysis performed so far on {\it EUVE} light curves (Audard et al. 1999; 2000; also
Kashyap et al. 2000) and
the new results presented here  suggest a distribution of flare energies
according to a power law with a {\bf steep index: $\alpha \approx 2-3$}. If the
flare rate distribution continues down to levels of average solar flares, then
the complete stellar corona could be heated solely by the energy released in
statistical flares. 

The results  suggest values of $\alpha \approx 2.1-2.5$ independent of whether only the
detected flares are investigated, or whether the complete emission is
simulated. This suggests that (i) the extrapolation  of the flare distribution
to undetected flare energy levels could add  large amounts of emission, and
(ii) that the complete statistical distribution of count rate levels is indeed
compatible with such a distribution. The two points require {\it a lower threshold}
for the power-law flare distribution to confine the emission to the observed level.
This has implicitly been taken into account as our simulated power-law distributions
contain flares only with energies above a pre-set lower threshold. The value of
the latter is determined by the normalization to the observation. 

Flares could then also explain why the temperatures of active stars become increasingly
hotter with `increasing activity': The higher flare rate keeps more (high-density)
plasma at hot temperatures, and thus the hotter plasma increasingly dominates the X-ray 
emission (G\"udel 1997; Audard et al. 2000).

\acknowledgments
We thank the CEA/{\it EUVE}, {\it VLA}, and 
{\it Beppo}SAX staff for their efforts to coordinate these observations.
M.~A. acknowledges support from the Swiss National Science Foundation, 
grants 2100-049343 and  2000-058827. The National Radio Astronomy Observatory 
is a facility of the National Science Foundation operated under cooperative 
agreement by Associated Universities, Inc. SRON is financially
supported by the Netherlands Organization for Scientific Research (NWO).

\end{document}